\documentclass[fl eqn,twoside]{article}
\date{}
\usepackage{amsmath}
\usepackage{amsfonts}
\usepackage{amssymb}
\usepackage{graphicx}
\date{\today}
\begin{document}

\title{{\bf{Path integral action of a particle with the generalized uncertainty principle and correspondence with noncommutativity }}}

\author{
{\bf {\normalsize Sunandan Gangopadhyay}
$^{a}$\thanks{ sunandan.gangopadhyay@bose.res.in, sunandan.gangopadhyay@gmail.com}},
{\bf {\normalsize Sukanta Bhattacharyya}$^{b}
$\thanks{sukanta706@gmail.com}}\\
$^{a}$ {\normalsize Department of Theoretical Sciences}\\
{\normalsize S.N. Bose National Centre for Basic Sciences}\\
{\normalsize JD Block, Sector III, Salt Lake, Kolkata 700106, India}\\
$^{b}$ {\normalsize Department of Physics, West Bengal State University, Barasat, Kolkata 700126, India}\\
}
\date{}
\maketitle
\begin{abstract}
\noindent The existence of an observer independent minimum length scale can lead to the modification of the Heisenberg uncertainty principle to the generalized uncertainty principle. This in turn would be responsible for the modification of the Hamiltonian describing a non-relativistic particle moving in the presence of an arbitrary potential. In this work we carry out a path integral formulation to compute the transition amplitude for this particle. The formalism yields the action of such a particle in an arbitrary potential. Interestingly, the action indicates that there is an upper bound to the velocity that a particle can have which depends on the generalized uncertainty principle parameter. We then compute explicitly the propagator of a free particle and particle moving in a harmonic oscilltor potential using the path integral representation of the transition amplitude. We observe that there exists a curious connection between the transition amplitude of the free particle in the generalized uncertainty priciple framework with the corresponding result in noncommutative space found from the path integral formulation in \cite{sgprl}. From the harmonic oscillator result for the transition amplitude, we calculate the ground state energy of the harmonic oscillator. The result shows that the ground state energy of the harmonic oscillator in the framework of the Heisenberg uncertainty principle gets augmented by the presence of the generalized uncertainty principle and also depends on the mass of the particle. We also demonstrate that the result agrees with that obatined using the operatorial approach.            

\end{abstract}
\vskip 1cm

\noindent The existence of an observer independent minimum length scale has been strongly proposed by various quantum gravity theories like loop quantum gravity \cite{rov,car}, string theory \cite{ama}, noncommutative geometry \cite{gir} to name a few. One of the consequences of this idea leads to a modification of the Heisenberg uncertainty principle (HUP) or equivalently, to a modification of the commutation relation between position and its conjugate momentum. This is well known in the literature as the generalized uncertainty principle (GUP). The possibility of the GUP has led to the investigation of various aspects in theoretical physics like black hole thermodynamics \cite{rj}-\cite{sg2}, quantum gravity corrections in various quantum systems such as particle in a box, Landau levels, simple harmonic oscillator \cite{sd}-\cite{sd3}. However, despite these studies not much attention have been paid to the path integral formulation for a nonrelativistic particle moving in an arbitrary potential in the GUP framework.
The literature contains an investigation of the free particle kernel in the presence of the GUP \cite{sp} but the analysis lacks in giving the transition amplitude in a path integral representation. Such a formulation would be necessary if not absolutely essential, to gain insight in our current understanding of quantum gravity phenomenology. For example, we can compare the action in this path integral formulation with that derived in a two dimensional noncommutative space \cite{sgprl}. The comparison should give us some insight about the nature of these two seemingly different quantum gravity phenomenological theories. Further, just as in the noncommutative case, a       closer analysis shows that the limit of infinite time slices is quite subtle in the present case also and requires some attention.

\noindent The theme of this paper is to derive the path integral representation and the classical action for a particle moving in an arbitrary potential in the presence of the GUP.  

\noindent To do this, we first write down the modified Poisson brackets between the position $\hat{q}_i$ 
and its conjugate momentum $\hat{p}_i$ incorporating the GUP. 
We use the simplest form of the GUP proposed in the literature \cite{kempf} 
\begin{eqnarray}
\Delta q_{i}\Delta p_{i}~\geq~\frac{\hbar}{2}\left[1+\beta(\Delta p^{2}+\langle
p\rangle^{2})+2\beta(\Delta p_{i}^{2}+<p_{i}>^{2})\right]~~~~i=1,2,3 \label{deltaxi}
\label{GUP}
\end{eqnarray}
where the dimension of the  GUP parameter $\beta$ is  $(momentum)^{-2}$.

\noindent The saturated form of the above inequality is equivalent to the following modified Heisenberg algebra \cite{kempf}
\begin{eqnarray}
[\hat q_{i}, \hat p_{j}]=i\hbar(\delta_{ij}+\beta\delta_{ij}\hat{p}^{2}+2\beta \hat{p}_{i}\hat{p}_{j}).
\label{GUP2}
\end{eqnarray}
It can be shown that the above commutation relation is readily obtained to first order in $\beta$ by defining the position and momentum operators as 
\begin{eqnarray}
\hat{q}_{i}=\hat{q}_{0i}~~~~~,~~~~~\hat{p}_{i}=\hat{p}_{0i}(1+\beta \hat{p}_{0}^{2}) 
\label{GUPR}
\end{eqnarray}
where $\hat{p}_{0}^{2}=\sum_{j=1}^{3}~\hat{p}_{0j}\hat{p}_{0j}$ and $\hat{q}_{0i},~\hat{p}_{0j}$ satisfy the usual canonical commutation relations $[\hat{q}_{0i}, \hat{p}_{0j}]=i\hbar\delta_{ij}$. 


\noindent With this background in place, the next step is to construct the Hamiltonian of a particle moving 
in an arbitrary potential $V(q)$ in the presence of the GUP. In the subsequent discussion we shall work in one spatial dimension. 
Using eq.(\ref{GUPR}) and taking the Hamiltonian to be of the form 
\begin{eqnarray} 
\hat H=\frac{\hat p^{2}}{2m}+V(\hat q)
\label{hamil} 
\end{eqnarray}
we find that it can be recast in the following form
\begin{eqnarray} 
\hat H=\frac{ \hat p_{0}^{2}}{2m}+\frac{\beta}{m}\hat p_{0}^{4}+V(\hat q) +\mathcal O(\beta^2).
\label{hamil11} 
\end{eqnarray}

\noindent We now proceed to construct the path integral representation of a particle with the GUP corrected Hamiltonian. 
In the rest of our discussion, we shall drop the suffix on $\hat{p}$ for notational convenience.
The following completeness relation is crucial for the construction of the path integral representation 
\begin{eqnarray}
\int_{-\infty}^{+\infty} dp~ |p\rangle \langle p|= \mathbf{1}~.
\label{cr}
\end{eqnarray}
The propagation kernel in general can be written as
\begin{eqnarray}
\langle q_{f}, t_{f}| q_{0}, t_{0} \rangle= \lim_{n \to \infty} \int_{-\infty}^{+\infty} \prod_{j=1}^{n} dq_{j} \langle q_{f},t_{f}| q_{n}, t_{n} \rangle \langle  q_{n}, t_{n}|.~.~.|q_{1},t_{1} \rangle \langle     q_{1},t_{1}|q_{0},t_{0} \rangle~.
\label{propafree}
\end{eqnarray}
\noindent Now we need to compute the propagator over a small segment in the above path integral. Using the Hamiltonian given by eq.(\ref{hamil11}) and the completeness relation (\ref{cr}), we can write the infinitesimal propagator as
\begin{eqnarray}
\langle q_{j+1},t_{j+1}|q_{j},t_{j}\rangle&=& \langle q_{j+1}|e^{-\frac{i}{\hbar}\hat H\tau}|q_{j}\rangle \nonumber\\
&=&\langle q_{j+1}|1-\frac{i}{\hbar}\hat H\tau+ \Theta(\tau^2)|q_{j} \rangle \nonumber\\&&
=\int_{-\infty}^{+\infty} \frac{dp_{j}}{2\pi\hbar}~~ e^{\frac{i}{\hbar} p_{j}(q_{j+1}-q_{j})} e^{-\frac{i}{\hbar}\tau(\frac{p_{j}^{2}}{2m}+\frac{\beta}{m}p_{j}^{4}+V(q))}+\Theta(\tau^2)
\label{propafree2}
\end{eqnarray}
where $p_j$ is the momentum following the ordinary Heisenberg uncertainty principle. Subtituting the above expression in eq.(\ref{propafree}), we obtain (apart from a constant factor)
\begin{eqnarray}
\langle q_{f}, t_{f}| q_{0}, t_{0}\rangle =\lim_{n \to \infty} \int_{-\infty}^{+\infty} \prod_{j=1}^{n} dq_{j} \prod_{j=0}^{n} dp_{j} ~exp \left(\frac{i}{\hbar} \sum_{j=0}^{n} \left[p_{j}(q_{j+1}-q_{j})-\tau(\frac{p_{j}^{2}}{2m}+\frac{\beta}{m}p_{j}^{4}+V(q_j))\right]\right).
\label{propa2}
\end{eqnarray}
In the $\tau \to 0$ limit, we finally have the phase-space form of the path integral 
\begin{eqnarray}
\langle q_{f}, t_{f}| q_{0}, t_{0}\rangle=\int \mathcal{D}q ~\mathcal{D}p~~ exp \left(\frac{i}{\hbar} \mathcal{A}\right)
\label{psprop}
\end{eqnarray} 
where $\mathcal{A}$ is the phase-space action given by 
\begin{eqnarray}
\mathcal{A}= \int_{t_0}^{t_f} dt \left[p\dot q- \left(\frac{p^{2}}{2m}+\frac{\beta}{m}p^{4}+V(q)\right)\right]~, ~t_f-t_0=T~.
\label{psaction}
\end{eqnarray}
\noindent Now we carry out the momentum integral in eq.(\ref{propafree2}). Evaluating this keeping terms upto $\mathcal{O}(\beta)$ yields
\begin{eqnarray}
\langle q_{j+1},t_{j+1}~|~q_{j},t_{j}\rangle&=&\sqrt{\frac{m}{2\pi i\hbar\tau}}
\left[1+\frac{3\beta i\hbar m}{\tau}
-\frac{6\beta m^{2}(q_{j+1}-q_{j})^{2}}{{\tau}^{2}}-\frac{i\beta m^{3}(q_{j+1}-q_{j})^{4}}
{\hbar{\tau}^{3}}\right]\nonumber\\&&
\times  exp\left({\frac{im(q_{j+1}-q_{j})^{2}}{2\hbar\tau}}+\frac{i}{\hbar}\tau V(q_j)\right)\nonumber\\
&=& \sqrt{\frac{m}{2\pi i\hbar\tau}}
\left[1+\frac{3\beta i\hbar m}{\tau}
-\frac{6\beta m^{2}(q_{j+1}-q_{j})^{2}}{{\tau}^{2}}\right]\nonumber\\&&
\times  exp\left({\frac{im(q_{j+1}-q_{j})^{2}}{2\hbar\tau}}+\frac{i}{\hbar}\tau V(q_j)-\frac{i\beta m^{3}(q_{j+1}-q_{j})^{4}}
{\hbar{\tau}^{3}}\right)+\mathcal{O}(\beta^2)
. \label{infpropafree3}
\end{eqnarray}
Using the above result in eq.(\ref{propa2}), we obtain (apart from a constant factor)
\begin{eqnarray}
\langle q_{f},t_{f} |q_{0}, t_{0} \rangle = \int_{-\infty}^{+\infty} \prod_{j=1}^{n} dq_{j}~~exp\left(\frac{i}{\hbar}\sum_{j=0}^{n} \tau \left[ \frac{m}{2}\left(\frac{q_{j+1}-q_{j}}{\tau}\right)^2\left\{1-2\beta m^2\left(\frac{q_{j+1}-q_{j}}{\tau}\right)^2\right\}-V(q_j)\right]\right)~
\label{propagafree1}
\end{eqnarray}

\noindent In the $\tau \to 0$ limit, we finally get the configuration space path integral representation of the particle moving in a potential $V(q)$ to be
\begin{eqnarray}
\langle q_{f},t_{f} |q_{0}, t_{0} \rangle =\tilde{N}(T, \beta) \int \mathcal{D}q~~e^{\frac{i}{\hbar}S}~
\label{propagafree3} 
\end{eqnarray} 
where the action of the particle moving in the presence of a potential $V(q)$ in the configuration space reads
\begin{eqnarray}
S=\int_{t_{0}}^{t_{f}} dt \left[\frac{m}{2} \dot q^2\left(1-2 \beta m^2 \dot q^2 \right) - V(q)\right]~.
\label{action}
\end{eqnarray}
Note that the constant $\tilde{N}(T, \beta)$ in eq.(\ref{propagafree3}) will be determined in the subsequent discussion. 

\noindent We make a few observations now. First, the above action shows that the Lagrangian of the particle moving in a potential $V(q)$ is given by this
\begin{eqnarray}
L=\left[\frac{m}{2} \dot q^2\left(1-2 \beta m^2 \dot q^2 \right) - V(q)\right]~.
\label{lag}
\end{eqnarray} 
This now leads to the Hamiltonian of the particle moving  in a potential $V(q)$ (by a Legendre transformation) to be
\begin{eqnarray}
H=\frac{p^{2}}{2m}+\frac{\beta}{m} p^{4}+V(q) +\mathcal O(\beta^2)~.
\label{hamil-lag}
\end{eqnarray}
This agrees with the form of the Hamiltonian operator (incorporating the effect of the GUP) with which we started our analysis.

\noindent With the general result in hand, we now study the simplest case which is that of the free particle ($V(q)$=0). To do this we first evaluate the classical action. This is easily done by writting down the classical equation of motion for the free particle. This reads
\begin{eqnarray}
m\left(1-12 \beta m^2 \dot q^2\right) \ddot q=0 \implies \ddot q=0 ~~~\textrm{or}~~~ \left(1-12 \beta m^2 \dot q^2\right)=0~.
\label{eqm}
\end{eqnarray}
Interestingly both possibilities yield $\ddot q=0$. Further there is an upper bound on the velocity of the free particle due to the presence of the GUP which can be easily seen from the action (\ref{action}). This is given by
\begin{eqnarray}
\dot{q}_{max}=\frac{1}{\sqrt{2 \beta m^2}}\implies \dot{p}_{max}=m\dot{q}_{max}=\frac{1}{\sqrt{2 \beta}}~.
\label{upb}
\end{eqnarray}
Interestingly, this bound on the momentum is consistent with the bound on the uncertainty in momentum \cite{sp} and thereby indicates the validity of our path integral formulation.

\noindent We now solve $\ddot q=0$ subject to the boundary conditions that at $t=t_0$, $q=q_0$; $t=t_f$, $q=q_f$ to get the classical trajectory of the free particle
\begin{eqnarray}
q_{c}(t)=q_0 + \frac{q_f-q_0}{T}t
\label{}
\end{eqnarray}
where $t_f-t_0=T$. Therefore the classical action for the free particle in the presence of the GUP takes the form
\begin{eqnarray}
S_{c}= \frac{m}{2 T} (q_{f}-q_{0})^2 \left[ 1- 2 \beta m^2 \left(\frac{q_{f}-q_{0}}{T}\right)^2\right]~.
\label{actionfree}
\end{eqnarray}
\noindent Using the above expression for the classical action in eq.(\ref{propagafree3}), we obtain 
\begin{eqnarray}
\langle q_{f},t_{f} |q_{0}, t_{0} \rangle = \tilde{N}(T,\beta)~e^{\frac{i m}{2 \hbar T} (q_{f}-q_{0})^2 \left[ 1- 2 \beta m^2 \left(\frac{q_{f}-q_{0}}{T}\right)^2\right]}~.
\label{fpir} 
\end{eqnarray}
Our next step is to evalute the constant $\tilde{N(T,\beta)}$ which contains the quantum fluctuations. To proceed we use the following identity
\begin{eqnarray}
\langle q_{f}, t_{f}|p\rangle= \int_{-\infty}^{+\infty}dq_0\left<q_{f},t_f|q_{0},t_{0}\right>\left<q_{0},t_{0}|p\right>~.
\label{ol}
\end{eqnarray}
Using eq.(\ref{fpir}) in eq.(\ref{ol}), setting $t_0=0$ and $t_f=T$ and using the overlaps
\begin{eqnarray}
\langle q_{0},0|p\rangle=\frac{1}{\sqrt{2 \pi \hbar}} exp({\frac{i}{\hbar}pq_0})~;~\langle q_f,T|p\rangle=\frac{1}{\sqrt{2\pi \hbar}}exp\left\{\frac{-i T}{\hbar}\left(\frac{p^{2}}{2m}+\frac{\beta}{m}p^{4}\right)\right\}~exp\left({\frac{i}{\hbar}pq_f}\right)
\label{ovl}
\end{eqnarray}
we get 
\begin{eqnarray}
\tilde{N}(T,\beta)=\sqrt{\frac{m}{2 \pi i \hbar T}}\left(1+\frac{3 i \beta \hbar m}{T}-\frac{6 \beta m^2(q_{f}-q_{0})^2}{T^2}\right).
\label{confree}
\end{eqnarray}
Eq.(\ref{fpir}) along with eq.(\ref{confree}) gives the propagation kernel for the free particle in the presence of the GUP
\begin{eqnarray}
\langle q_{f},t_{f} |q_{0}, t_{0} \rangle =\sqrt{\frac{m}{2 \pi i \hbar T}}\left(1+\frac{3 i \beta \hbar m}{T}-\frac{6 \beta m^2(q_{f}-q_{0})^2}{T^2}\right) ~e^{\frac{i m}{2 \hbar T} (q_{f}-q_{0})^2 \left[ 1- 2 \beta m^2 \left(\frac{q_{f}-q_{0}}{T}\right)^2\right]}~.
\label{full}
\end{eqnarray}
We now compare the above form of the transition amplitude with that of a particle moving in an arbitrary potential in two dimensional noncommutative plane. This reads \cite{sgprl}
\begin{eqnarray}
(z_f, t_f|z_0, t_0)=\sqrt{\frac{m}{2\pi (i\hbar T+ m\theta)}}\exp\left[-\frac{m}{2(i\hbar T +m\theta)}(\vec{x}_f -\vec{x}_0)^2\right]
\label{nctr}
\end{eqnarray}
where $z=\frac{1}{\sqrt{2\theta}}(x+iy)$ is a dimensionless complex number and $\theta$ is the noncommutative parameter arising from $[\hat x, \hat y]=i\theta$, $\theta$ being a real positive number. The above result can be recast in the form
\begin{eqnarray}
(z_f, t_f|z_0, t_0)=\sqrt{\frac{m}{2\pi i\hbar T}}\left[1+\frac{im\theta}{2\hbar T}-\frac{m^2 \theta}{2\hbar^2}\left(\frac{\vec{x}_f -\vec{x}_0}{T}\right)^2\right]\exp\left[-\frac{m}{2(i\hbar T +m\theta)}(\vec{x}_f -\vec{x}_0)^2\right].
\label{nctr1}
\end{eqnarray}
We now observe that there is a striking similarity between the forms of the factors in front of the exponential terms in eq.(\ref{full}) and eq.(\ref{nctr1}). The difference between these two results is in the form of the exponents 
which in the GUP case involves a term $(\frac{q_f -q_0}{T})^4$ in contrast to the noncommutative case which does not involve such a term. The reason for this difference lies in the fact that the action for the GUP case (\ref{action}) involves a $\dot{q}^4$ term whereas the action for the particle in the noncommutative plane has the form \cite{sgprl}
\begin{eqnarray}
S=\int_{t_0}^{t_f}dt\left[\frac{\theta}{2}\dot{\bar{z}}(t)\left(\frac{1}{2m}-\frac{i\theta}{2\hbar}\partial_{t}\right)^{-1}\dot{z}(t)-V(\bar{z}(t), z(t))\right].
\label{ncaction}
\end{eqnarray}
The above action is non-local in time which in the free particle case yields eq.(\ref{nctr}).

\noindent Now we extend our analysis upto $\mathcal{O}(\beta^2)$ in the GUP parameter $\beta$. The Hamiltonian of the free particle upto $\mathcal{O}(\beta^2)$ reads
\begin{eqnarray}
\hat H= \frac{\hat{p}^2}{2m} + \frac{\beta \hat{p}^2}{m} + \frac{\beta^2 \hat{p}^6}{2 m}~.
\label{hgoredr}
\end{eqnarray} 
\noindent Following the same procedure as earlier, we get the action for the free particle upto $\mathcal{O}(\beta^2)$ to be
\begin{eqnarray}
S=\int_{t_{0}}^{t_{f}} dt \left[\frac{m}{2} \dot q^2\left(1-2 \beta m^2 \dot q^2  -15 \beta^2 m^4\dot q^4\right)\right]~.
\label{action2}
\end{eqnarray}
From the action we observe that
\begin{eqnarray}
\left(1-2 \beta m^2 \dot q^2  -15 \beta^2 m^4\dot q^4\right)\geq 0~.
\label{}
\end{eqnarray}
\noindent We now proceed to the case of a particle moving in the presence of a harmonic oscillator potential.
\noindent In this case the action of the particle reads 
\begin{eqnarray}
S=\int_{0}^{T} dt \left[\frac{m}{2} \dot q^2\left(1-2 \beta m^2 \dot q^2\right)-\frac{1}{2} m \omega^2 q^2 \right]
\label{actionho}
\end{eqnarray}
where we have set $t_0=0$ and $t_f= T$.
The classical equation of motion following from the above action reads
\begin{eqnarray}
\ddot q(t)-12 \beta m^2 \dot q^2(t)\ddot q(t) + \omega^2 q(t)=0~.
\label{eqmho}
\end{eqnarray}  
\noindent Solving the equation of motion, the classical trajectory of the harmonic oscillator in the presence of the GUP  upto first order in the GUP parameter $\beta$ takes the form 
\begin{eqnarray}
q_{cl}(t)&=& A \cos ( \omega t)+B \sin (
\omega t)+\beta  \left[F \cos (\omega t) + H \sin ( \omega t) - 12 m^2 \left\{\frac{1}{8} t \omega^3  A(A^2+ B^2) \sin ( \omega t) \right.\right.\nonumber\\ && \left.\left.
+\frac{1}{32} \omega^2  A(A^2-3 B^2) \cos (3 \omega t)
-\frac{1}{8} t \omega^3  B(A^2+B^2) \cos ( \omega t)+\frac{1}{32} \omega^2  B (3 A^2-B^2) \sin (3 \omega t)\right\}\right]
\label{cthar}
\end{eqnarray}
where $A$,$B$,$F$ and $H$ are constants. Using the initial conditions that at $t=0$, $q=q_0$; $t=T$, $q=q_f$, the constants get determined and are given by
\begin{eqnarray}
A&=&q_0 \nonumber\\ 
B&=&[q_f-q_0 \cos ( \omega T)]\csc (\omega T)\nonumber\\
F&=&\frac{3}{8} m^2 \omega^2 \left(A^3-3 A B^2\right) \nonumber\\ 
H&=&12 m^2 \omega^2 \csc (\omega T) \left[\frac{1}{8}  \omega T \left(A^3+A B^2\right) \sin (\omega T)+\frac{1}{32} \left(A^3-3 A B^2\right) \cos (3 \omega T)+\frac{1}{32} \left(3 A^2 B-B^3\right) \sin (3 \omega T)\right.\nonumber\\ && \left.
-\frac{1}{8} \omega T \left(A^2 B+B^3\right) \cos (\omega T)\right]-\frac{3}{8} m^2 \omega^2 \left(A^3-3 A B^2\right) \cot (\omega T) ~.
\label{conshar}
\end{eqnarray}
Using eqs.(\ref{cthar},~\ref{conshar}) in eq.(\ref{actionho}), we obtain the classical action for the GUP corrected harmonic oscillator
\begin{eqnarray}
S_{c}= S_{c}(\beta=0)+S_{c}(\beta)
\label{hofa}
\end{eqnarray}
where $S_{c}(\beta)$ are the terms involving first order in the GUP parameter $\beta$. The forms of  $S_{c}(\beta=0)$ and $S_{c}(\beta)$ read
\begin{eqnarray}
S_{c}(\beta=0)=\frac{1}{2} m w \csc(wT) \left[\left(q_0^2+q_f^2\right) \cos (w T)-2 q_0 q_f\right]
\label{hoao}
\end{eqnarray}
\begin{eqnarray}
S_{c}(\beta)&=&-\frac{1}{128}\beta  m^3 w \csc ^5( w T) \bigg[\left(w^2+3\right) \left(q_0^4+q_f^4\right) \cos (5 w T)-4 q_0 q_f \left(q_0^2+q_f^2\right) \{w^2 \left(60 T^2
(w^2+1)-11\right)-33\} \nonumber\\&&  -16 q_0 q_f \left(q_0^2+q_f^2\right) \{w^2 \left(9 T^2 (w^2+1)+2\right)+6\} \cos (2 w T)-24 q_0 q_f(5 w^2+1) \left(q_0^2+q_f^2\right)  w  T \sin (2 w T)\nonumber\\&& -12 q_0 q_f (w^2+1) \left(q_0^2+q_f^2\right) w T  \sin (4 w T)-12 q_0 q_f (w^2+3) \left(q_0^2+q_f^2\right) \cos (4 w T)\nonumber\\&& +4 \left\{2(q_0^4+ q_f^4) \left( w^2 \left(12 T^2 (w^2+1)-1\right)-3\right)+3 q_0^2 q_f^2 \left(w^2 \left(44 T^2
(w^2+1)-5\right)-15\right)\right\}  \cos (w T) \nonumber\\&&  + \left\{7 (q_0^4+ q_f^4) (w^2+3)+12 q_0^2 q_f^2 \left(w^2 \left(4 T^2 (w^2+1)+5\right)+15\right)\right\}\cos (3 w T)  \nonumber\\&& +24  \left\{-2( q_0^4+ q_f^4)+3 q_0^2 q_f^2
(3 w^2+1)\right\} w T \sin (w T) +24 \left\{(q_0^4 +q_f^4 )(w^2+1) \right.\nonumber\\&& \left.+q_0^2 q_f^2 (3 w^2+1)\right\} w T  \sin (3 w T)\bigg]~.
\label{hoag}
\end{eqnarray}

\noindent  It is reassuring to note that we recover the free particle classical action (\ref{actionfree}) in the limit $\omega \to 0$. The propagator for the harmonic oscillator therefore takes the form
\begin{eqnarray}
\langle q_{f},t_{f}|q_{0},t_{0}\rangle= \sqrt{\frac{m \omega}{2 \pi i \hbar \sin(\omega T)}}\left[1+\frac{3\beta i \hbar m}{T}-6 \beta m^2 \left(\frac{q_f-q_0}{T}\right)^2-\frac{3}{4}\beta m \hbar \omega^2 T \cot(\omega T)\right] ~ e^{\frac{i}{\hbar} S_{c}}~.
\label{fprho}
\end{eqnarray}

\noindent From this we can evaluate the ground state energy of the harmonic oscillator in the presence of the GUP. This reads 
\begin{eqnarray}
E_0=\frac{1}{2}\hbar \omega\left( 1 + \frac{3}{2} \beta m \hbar \omega\right)~.
\label{gseho}
\end{eqnarray}
\noindent  The result shows that the ground state energy of the harmonic oscillator in the framework of the Heisenberg uncertainty principle gets modified by the presence of the GUP and also depends on the mass of the particle. The result could in principle be of great importance since it captures the effects of quantum gravity in the zero point energy of the harmonic oscillator which in turn could have profound implications in quantum field theory and its predictions, for example the Casimir effect. 

\noindent To cross-check the correctness of our result, we shall now calculate the ground state energy of the harmonic oscillator in the presence of the GUP in the operatorial approach. To do this we first break the Hamiltonian (\ref{hamil11}) as
\begin{eqnarray}
\hat H = \hat H_0+ \hat H_1
\label{break}
\end{eqnarray} 
where $\hat H_0$ and $\hat H_1$ are given by
\begin{eqnarray}
\hat H_0&=& \frac{\hat p^2}{2m}+\frac{1}{2} m w^2 \hat q^2 \nonumber\\
\hat H_1&=& \frac{\beta}{m} \hat p^4~.
\label{bk1}
\end{eqnarray}
\noindent  We now calculate the effect of the GUP by using perturbation theory. To do this we first recast $\hat H_1$ as
\begin{eqnarray}
\hat H_1 &=& 4 \beta m \left[\hat H_0^2+\frac{m^2 w^4}{4} \hat q^4+\frac{i \hbar w^2}{2}\left(2 \hat q \hat p-i\hbar\right)- m w^2 \hat q^2 \hat H_0\right]~.
\label{bk2}
\end{eqnarray}
\noindent With this form of the perturbed Hamiltonian, we now proceed to calculate the first order correction to the ground state energy of the harmonic oscillator. This can be done by following the standard quantum mechanical prescription and gives
\begin{eqnarray}
\langle \psi_0|\hat H_1|\psi_0 \rangle &=& 4 \beta m \left[\langle \psi_0|\hat H_0^2|\psi_0 \rangle +\frac{m^2 w^4}{4}\langle \psi_0| \hat q^4|\psi_0 \rangle +\frac{i \hbar w^2}{2}\left(\langle \psi_0|2 \hat q \hat p|\psi_0 \rangle-i\hbar\langle \psi_0|\psi_0 \rangle \right) \right. \nonumber\\&&
 \left. - m w^2 \langle \psi_0|\hat q^2 \hat H_0|\psi_0 \rangle \right]
\label{bk3}
\end{eqnarray}
where $|\psi_0 \rangle $ is the ground state of the unperturbed harmonic oscillator.  

\noindent Calculating the above matrix elements, we obtain the first order correction to the ground state energy of the harmonic oscillator to be
\begin{eqnarray}
\langle \psi_0|\hat H_1|\psi_0 \rangle &=& \frac{3}{4} \beta m w^2 \hbar^2
\label{bk4}
\end{eqnarray} 
which agrees with the result obtained from the path integral approach as can be seen from eq.({\ref{gseho}}).

\noindent We now summarize our findings. In this paper we have formulated the path integral 
representation of the transition amplitude of a particle moving in an arbitrary potential 
in the framework of the generalized uncertainty principle. From the path integral, we have
obtained the action of the particle. This is the central
result of this work. From the form of the action we observe that there is an upper bound to the velocity of the particle which depends on the generalized uncertainty principle parameter. We note that this upper bound on the velocity is consistent with the upper bound in the momentum uncertainty observed earlier in the literature. We then extend our analysis to second order in the generalized uncertainty principle parameter for the free particle case.
We then calculate the propagator of a free particle and particle moving in a harmonic oscilltor potential using the path integral representation of the transition amplitude. We observe that there is a surprising connection between the transition amplitude of the free particle in the generalized uncertainty priciple framework with the corresponding result in noncommutative space found from the path integral formulation in \cite{sgprl}. From the harmonic oscillator result for the transition amplitude, we then calculate the ground state energy of the harmonic oscillator. The result shows that in the framework of the generalized uncertainty principle the ground state energy of the harmonic oscillator gets augmented by an extra amount which depends on the generalized uncertainty priciple parameter. 
We speculate that the result can be of great significance since it captures the effects of quantum gravity in the zero point energy of the harmonic oscillator which could be responsible for having profound implications in quantum field theory and its predictions. We finally demonstrate that the result agrees with that obtained using the operatorial approach.            

\section*{Acknowledement}
S.G. acknowledges the support by DST SERB under Start Up Research Grant (Young Scientist), File No.YSS/2014/000180. SG also acknowledges IUCAA, Pune for the Visiting Associateship. SB would like to thank the Govt. of West Bengal for financial support.


\end{document}